\documentstyle[11pt]{article}
\begin{document}
\newcommand{\Si}{\Sigma}
\newcommand{\tr}{{\rm tr}}
\newcommand{\ad}{{\rm ad}}
\newcommand{\Ad}{{\rm Ad}}
\newcommand{\ti}[1]{\tilde{#1}}
\newcommand{\om}{\omega}
\newcommand{\Om}{\Omega}
\newcommand{\de}{\delta}
\newcommand{\al}{\alpha}
\newcommand{\te}{\theta}
\newcommand{\vth}{\vartheta}
\newcommand{\be}{\beta}
\newcommand{\la}{\lambda}
\newcommand{\La}{\Lambda}
\newcommand{\D}{\Delta}
\newcommand{\ve}{\varepsilon}
\newcommand{\ep}{\epsilon}
\newcommand{\vf}{\varphi}
\newcommand{\G}{\Gamma}
\newcommand{\ka}{\kappa}
\newcommand{\ip}{\hat{\upsilon}}
\newcommand{\Ip}{\hat{\Upsilon}}
\newcommand{\ga}{\gamma}
\newcommand{\ze}{\zeta}
\newcommand{\si}{\sigma}
\def\bfa{{\bf a}}
\def\bfb{{\bf b}}
\def\bfc{{\bf c}}
\def\bfd{{\bf d}}
\def\bfm{{\bf m}}
\def\bfn{{\bf n}}
\def\bfp{{\bf p}}
\def\bfu{{\bf u}}
\def\bfv{{\bf v}}
\def\bft{{\bf t}}
\def\bfx{{\bf x}}
\newcommand{\li}{\lim_{n\rightarrow \infty}}
\newcommand{\mat}[4]{\left(\begin{array}{cc}{#1}&{#2}\\{#3}&{#4}
\end{array}\right)}
\newcommand{\beq}[1]{\begin{equation}\label{#1}}
\newcommand{\eq}{\end{equation}}
\newcommand{\beqn}[1]{\begin{eqnarray}\label{#1}}
\newcommand{\eqn}{\end{eqnarray}}
\newcommand{\p}{\partial}
\newcommand{\di}{{\rm diag}}
\newcommand{\oh}{\frac{1}{2}}
\newcommand{\su}{su(2)}
\newcommand{\uo}{{\bf u_1}}
\newcommand{\Sl}{{\rm sl}(2,{\bf C})}
\newcommand{\SL}{{\rm SL}(2,{\bf C})}
\def\SU{{\rm SU}(2)}
\def\GLN{{\rm GL}(N,{\bf C})}
\newcommand{\PSL}{{\rm PSL}_2({\bf Z})}
\def\f1#1{\frac{1}{#1}}
\newcommand{\rar}{\rightarrow}
\newcommand{\upar}{\uparrow}
\newcommand{\sm}{\setminus}
\newcommand{\ms}{\mapsto}
\newcommand{\bp}{\bar{\partial}}
\newcommand{\bz}{\bar{z}}
\newcommand{\bA}{\bar{A}}
\newcommand{\bL}{\bar{L}}
\newcommand{\ot}{\otimes}

%\def\theequation{\thesubsection.\arabic{equation}}% the equation
               % number now does not include the section number;
               % \setcounter{equation}{0} should be put after every
               % \section{} command!!!
\newcommand{\sect}[1]{\setcounter{equation}{0}\section{#1}}
\renewcommand{\theequation}{\thesection.\arabic{equation}}

\newtheorem{predl}{Proposition}[section]
\newtheorem{defi}{Definition}[section]
\newtheorem{rem}{Remark}[section]
\newtheorem{cor}{Corollary}[section]
\newtheorem{lem}{Lemma}[section]
\newtheorem{theor}{Theorem}[section]

\vspace{0.3in}
\begin{flushright}
 ITEP-TH-2/99\\
\end{flushright}
\vspace{10mm}
\begin{center}
{\Large\bf
Painlev\'{e} type equations\\
and Hitchin systems}\footnote
{Contribution in the Proceedings 
"Integrability: the Seiberg-Witten and Whitham Equations",
Edinburgh, September 1998}

\vspace{5mm}

M.A.Olshanetsky
\\

{\sf Max-Plank-Institut f\"{u}r Matematik, Bonn}\\
{\sf Institut of Theoretical and Experimental Physics, 
Moscow, Russia,}\\

{\em e-mail olshanet@mpim-bonn.mpg.de}\\
{\em e-mail olshanet@heron.itep.ru}\\

\vspace{5mm}
\end{center}
\begin{abstract}
In this survey we present the interpretation of isomondromy
preserving equations on Riemann surfaces with marked points 
as reduced Hamiltonian systems. The upstairs space is the space of 
smooth connections of GL(N) bundles with simple poles in the marked
points. We discuss relations of these equations with 
the Whitham quantization of the Hitchin systems and with the
classical limit of the Knizhnik-Zamolodchikov-Bernard equations.
The main example is the one-parameter family of Painlev\'{e} VI
equation and its multicomponent generalization.

\end{abstract}

\vspace{0.15in}
\bigskip
\title
\maketitle
\date{}

\section {Introduction}
\setcounter{equation}{0}
The famous Painlev\'{e} VI
equation depends on four free parameters
$(PVI_{\al,\be,\ga,\de})$ and has the
form
$$
\frac{d^2X}{dt^2}=\frac{1}{2}\left(\frac{1}{X}+\frac{1}{X-1}+
\frac{1}{X-t}\right)
\left(\frac{dX}{dt}\right)^2-
\left(\frac{1}{t}+\frac{1}{t-1}+\frac{1}{X-t}\right)\frac{dX}{dt}+
$$
\beq{I.1}
+\frac{X(X-1)(X-t)}{t^2(t-1)^2}\left(\al+\be\frac{t}{X^2}+
\ga\frac{t-1}{(X-1)^2}
+\de\frac{t(t-1)}{(X-t)^2}\right).
\eq
It was discovered by B.Gambier \cite{Gam} in 1906.
He accomplished the Painlev\'{e} 
classification program
of the second order differential equations whose solutions have not
movable critical points. This equation and its degenerations $PV-PI$ have
a lot of applications in classical and quantum integrable systems 
(see, for example \cite{LW}), topological field theories \cite{Du}, 
general relativity \cite{Hi2,KN}, and 
in the Seiberg-Witten theory \cite{HP}.
In this paper we discuss two important and interrelated  
aspects of PVI:\\
$\bullet$ PVI and isomonodromic deformations of linear differential equations;\\
$\bullet$ The Hamiltonian structure of PVI.\\
The derivation of PVI equation as the preserving monodromy condition
was given by R.Fuchs \cite{Fu}, while the Hamiltonian structure of PVI 
was introduced by J.Malmquist \cite{Mal}.

We incorporate the one-parameter family 
PVI$_{\frac{\nu^2}{4},-\frac{\nu^2}{4},
\frac{\nu^2}{4},\f1{2}-\frac{\nu^2}{4}}=$PVI$_\nu$ in a wide class 
of nonlinear equations. They preserve monodromies of
systems of linear equations on Riemann curves with marked points
when the complex structures of curves are changed.
These systems come
from the flatness condition of vector bundles on the curves. We restrict
 ourself by considerations of smooth connections with 
simple poles only,
and therefore don't include the Stokes phenomena. 
In such general form the isomondromy preserving equations were considered
in \cite{Iw}. Our investigation of these systems is inspired by methods
developed in classical and quantum integrable systems.
In general all the systems can be derived using three 
different constructions:\\
A)The symplectic reduction procedure from free infinite dimensional
theory. This approach is very similar to the derivation of the 
Hitchin integrable systems \cite{H};\\
B)The Whitham quantization of the Hitchin systems \cite{Kr};\\
C)Classical limit of the Knizhnik-Zamolodchikov-Bernard (KZB) equations
\cite{KZ,Be}.\\
We discuss these constructions separately and then demonstrate 
their application on the multicomponent generalization of PVI$_\nu$.
The presentation is based for the most part on a previously
published paper \cite{LO}. First, in Sect.2 we consider the elliptic 
form of PVI. In this form the relations of PVI with the Hitchin systems 
and KZB equations become transparent. Then we discuss these three approaches
to the isomonodromic deformations. In Sect.6 the  multicomponent generalization
of PVI is described. Finally, we discussed some open problems related to
  PVI and its generalizations.

\section{Elliptic form of PVI}
\setcounter{equation}{0}
{\bf 1. Elliptization procedure}\\
Soon after discovering of PVI (\ref{I.1})  by Gambier, 
Painlev\`{e} presented it in terms of the Weierstrass
 elliptic functions \cite{Pe}. This paper was almost
forgotten for ninety years and the elliptic form was rediscovered
recently in \cite{Ma,BB}. We follow the derivation
 presented in  \cite{Ma}, where the Hamiltonian form and symmetrices
of PVI in terms of elliptic functions are treated.

Consider the family of elliptic curves
\beq{ef}
E_{\tau}={\bf C}/({\bf Z}+{\bf Z}\tau)
\eq
 where $\tau\in H=\{Im\tau>0\}$
Let $\wp(u|\tau)$ be the Weierstrass function
\beq{1.1}
\wp(u|=\f1{u^2}+\sum '
\left(\f1{(u+m\om_1+n\om_2)^2}-
\f1{(m\om_1+n\om_2)^2}
\right),~(\tau=\frac{\om_2}{\om_1}).
\eq
In the most part of the paper we put $\om_1=1$ and
$\wp(u|\tau)=\wp(u|1,\om_2)$.
$\wp(u|\tau)$ uniformize the elliptic curve
\beq{1.2}
\wp_u(u|\tau)=4(\wp(u|\tau)-e_1(\tau))(\wp(u|\tau)-e_2(\tau))
(\wp(u|\tau)-e_3(\tau)),
\eq
$$
e_i=\wp\left(\frac{T_i}{2}|\tau\right),~~
(T_0,\ldots,T_3)=(0,1,\tau,1+\tau).
$$
We consider two kind of transformations. First one is the
lattice action 
\beq{1.3}
u\rar u+m+n\tau, ~\tau\rar\tau.
\eq
It leaves $\wp(u|\tau)$ and $\wp_u(u|\tau)$ invariant. The second is
the modular transformation by $\PSL$
\beq{2.4}
\wp\left(
\frac{u}{c\tau+d}|
\frac{a\tau+b}{c\tau+d}\right)=(c\tau+d)^2\wp(u|\tau),~~
\wp_u\left(
\frac{u}{c\tau+d}|
\frac{a\tau+b}{c\tau+d}\right)=(c\tau+d)^3\wp_u(u|\tau).
\eq

 Now consider another family of
elliptic curves $E_t\rar B,~~Y^2=X(X-1)(X-t)$
parameterized by $B=\{t\in{\bf P}^1\setminus (0,1,\infty)\}$.
There exists the morphism $\{E_\tau\}\rar\{E_t\}$ defined as
\beq{I.2}
(u,\tau)\rar 
\left(X=\frac{\wp(u|\tau)-e_1}{e_2-e_1},
Y=\frac{\wp_u(u|\tau)}{e_2-e_1},
t=\frac{e_3-e_1}{e_2-e_1}\right).
\eq
\begin{theor}
In terms of $(u,\tau)$ $PVI_{\al,\be,\ga,\de}$ takes the form
\beq{I.3}
\frac{d^2u}{d\tau^2}=\p_uU(u|\tau),~~
U(u|\tau)=
\frac{1}{(2\pi i)^2}\sum_{j=0}^3\al_j\wp(u+\frac{T_j}{2}|\tau),
\eq
$$
(\al_0,\ldots,\al_3)=(\al,-\be,\ga,\f1{2}-\de).
$$
\end{theor}

The proof of the equivalence of (\ref{I.1}) and (\ref{I.3}) is based
 on the Picard-Fuchs equation on elliptic curves. 
The Picard-Fuchs operator
$$
L_t=t(t-1)\frac{\p^2}{\p t^2}+(1-2t)\frac{\p}{\p t}-
\frac{1}{4}
$$
acting on the holomorphic differential $\om=(d_{E/B}x)/y$ yields
the exact differential \\ $\oh d_{E/B}\frac{y}{(x-t)^2}$.
The Picard-Fuchs equation just means that periods of
$d_{E/B}x/y$ are annihilated by $L_t$. Using the Picard-Fuchs
operator Fuchs  proved that PVI (\ref{I.1}) is equivalent to 
the following equation
\beq{I.5}
t(1-t)L_t\int_\infty^Xd_{E/B}x/y=\left(\al+\be\frac{t}{X^2}+
\ga\frac{t-1}{(X-1)^2}+(\de-\oh)\frac{t(t-1)}{(X-t)^2}\right)Y,
\eq
The equivalence of (\ref{I.1}) and (\ref{I.5}) follows from
the following equality
$$
t(1-t)L_t\int_\infty^Xd_{E/B}x/y= \oh\frac{t(t-1)Y}{(X-t)^2}
+
\frac{d^2X}{dt^2}-\frac{1}{2}\left(\frac{1}{X}+\frac{1}{X-1}+\frac{1}{X-t}
\right)
\left(\frac{dX}{dt}\right)^2+
$$
$$
\left(\frac{1}{t}+\frac{1}{t-1}+\frac{1}{X-t}\right)\frac{dX}{dt},~~
Y^2=X(X-1)(X-t).
$$
The proof is straightforward.

 Thus, PVI can be written in the form of the so-called 
$\mu$-equation \cite{Ma}
\beq{mu}
L_t\int_\infty^X\om=s_{(\al,\be,\ga,\de)}(X),
\eq
where the right hand side is a special section of the bundle $E_t$.
It can be fixed by the symmetries of the equation.

Under the morphism (\ref{I.2}) the holomorphic differential $d_{E/H}z$ on 
$E_\tau$  is transformed in $d_{E/B}x/y$, and 
 $\frac{d^2}{d\tau^2}$ in $L_t$.
More exactly, the left hand side of 
(\ref{mu}) takes the form
$$
\frac{2\pi^2}{(e_1-e_2)(e_1-e_3)(e_2-e_1)^\oh}\frac{d^2}{d\tau^2}
\int_0^{u(\tau)}dz.
$$
Taking into account that
$$
Y=\oh(e_2-e_1)^{-3/2}\wp_u(u,\tau)
$$ 
we come finally to (\ref{I.3}).

\bigskip
\noindent

{\bf 2. Hamiltonian structure}\\

 The hamiltonian form of (\ref{I.3}) is defined
by the standard symplectic form
\beq{I.6}
\om^{(0)}=\de v\de u,
\eq
and the Hamiltonian
\beq{I.7}
H=\frac{v^2}{2}-U(u|\tau).
\eq
Consider the bundle ${\cal P}$ over the moduli space ${\cal M}=H/\PSL$ 
with the symplectic fibers parameterized by the  local
coordinates $(v,u)$. It plays role of the extended phase space for the
 non-autonomous hamiltonian system (\ref{I.6}),(\ref{I.7}).
The equation of motion (\ref{I.3}) can be derived from the action ${\cal F}$
 on ${\cal P}$
\beq{I.5a}
\de {\cal F}=v\de u-H\de\tau.
\eq

 The symmetries of the non-autonomous hamiltonian systems are determined
by the invariance of 
the two-form $\om$ on ${\cal P}$
\beq{I.5b}
\om=\om^{(0)}-\de H\de\tau=\de v\de u-\de H\de\tau.
\eq
It follows from (\ref{1.1}) and (\ref{2.4}) that the symmetry group is
 the semi-direct product of ${\bf Z}+{\bf Z}\tau$ and the group 
$\G(2)\subset\PSL$. We consider a simplified version of this action in
 Sect.7 in detail.

{\bf 3. Calogero-Inozemtsev equation and PVI}\\

Let us introduce the new parameter $\ka$ and instead of (\ref{I.5b}) 
consider
\beq{I.6a}
\om=\om_0-\f1{\ka}\de H\de\tau.
\eq
It can be achieved by the rescaling the dynamical variables $(v,u)$ and 
periods $\om_1,\om_2$ 
$
v\rar\ka^{-\oh},~u\rar\ka^{\oh},~\om_1\rar\ka^{\oh},~\om_1\rar\ka^{\oh}.
$
Then, (\ref{I.3}) takes the form
\beq{I.3a}
\ka^2\frac{d^2u}{d\tau^2}=-\p_uU(u|\tau).
\eq
Put $\tau=\tau_0+\ka t^H$ and consider the system in the limit
 $\ka\rar 0$.
We come to the equation
\beq{I.8}
\frac{d^2u}{(dt^H)^2}=-\p_uU(u|\tau_0)
\eq
corresponding to the autonomous Hamiltonian system with the 
time-independent potential $U(u|\tau_0)$. 
It is just the rank one elliptic Calogero-Inozemtsev equation
 $(CI_{\al,\be,\ga,\de})$ \cite{Ca,In}. The potential
$U(u|\tau_0)$  was considered first by Darboux \cite{Da}. It 
 arises also in the soliton theory \cite{TV}.
Thus, we have in this limit
\beq{I.7a}
PVI_{\al,\be,\ga,\de}\stackrel{\ka\rar 0}
{\longrightarrow}CI_{\al,\be,\ga,\de}.
\eq

There is the inverse procedure (the Whitham quantization) that
allows to construct approximations of non-autonomous systems starting 
from integrable autonomous systems. It will be discussed in Sect.4.

Inozemtsev considered degenerations of $U(u|\tau_0)$ playing with
the coupling constants, the periods, and $u$. In this way he obtained
the trigonometric, rational and exponential interactions. Presumably,
they describe the degenerations of PVI to PV-PI in terms
of degenerations of elliptic functions.
Here is one of his potentials:
$$
\al_0\f1{\sinh^2u}+\al_1\f1{\sinh^22u}+\al_2\exp u+\al_3\exp 2u.
$$
In what follows we consider only the subfamily PVI$_\nu$ corresponding
to $\al_j=\nu^2$.

\section {Isomonodromic deformations}

\setcounter{equation}{0}
Here we describe the monodromy preserving equations as reduced
Hamiltonian systems. The original phase space is infinite-dimensional
and almost all degrees of freedom are killed by the symplectic reduction.
Our approach differs from \cite{Ha}, where  PI-PVI equations are treated
as a result of symplectic reduction from finite-dimensional space.

{\bf 1. Hamiltonian approach}\\
Let $\Si_g$ be a Riemann curve of genus $g$. Consider the  space
$FBun_{\Si,G}$ of flat vector bundle $V_G$, where $G=\GLN$ with smooth
connection ${\cal A}$. The flatness
means that its curvature vanishes
\beq{2.1}
F_{{\cal A}}=d{\cal A}+\oh[{\cal A},{\cal A}]=0 .
\eq
Let us fix the complex structure on $\Si_g$. Then  
for ${\cal A}=(A,\bA)$ we have locally 
a consistent system of matrix differential equations 
$$
(\p+A)\Psi=0,
$$
$$ 
(\bp+\bA)\Psi=0.
$$
We modify this system in the following way. First, introduce formally
a parameter $\ka\in {\bf R}$ ({\sl the level}) and consider the operator
$\ka\p$ instead of $\p$ in the first equation. 
Let $\mu$ be a Beltrami differential on 
$\Si_g$ ($\mu\in\Om^{(-1,1)}(\Si_g)$).
 It means that in local coordinates \\
$\mu=\mu(z,\bz)\frac{\p}{\p z}\otimes d\bz$.
It allows to deform the complex structure on $\Si_g$ such that
the new complex coordinates are
$$
w=z-\ep(z,\bz),~~\bar{w}=\bz,~~~
\mu(z,\bz)=\frac{\bp\ep(z,\bz)}{1-\p\ep(z,\bz)}.
$$
The holomorphic operator $\p_{\bar{w}}= \bp+\mu\p$ is 
 annihilates the one-form $dw$ likewise
$\bp$ annihilates $dz$. 
We do not touch the anti-holomorphic operator $\ka\p$. 
In the new coordinates  (\ref{2.1}) takes the form
\beq{fl2}
F_{\cal A}=(\bp +\p\mu)A-\ka\p\bA+[\bA,A]=0.
\eq
Thus, we come to the system
\beq{ls1}
(\ka\p+A)\Psi=0,
\eq
\beq{ls2}
(\bp+\mu\p+\bA)\Psi=0.
\eq
Represent the Beltrami differential as
$
\mu=\sum_{a=1}^lt_a\mu_a^0,
$
 where $\mu^0_1,\ldots,\mu^0_l$ is the basis in the tangent space to
the moduli space ${\cal M}_g$ of complex structures on $\Si_g$,
($l=\dim{\cal M}_g=3g-3$, for $g>1$, $l=1$, for $g=1$). In other words, 
$\bft=(t_1,\ldots,t_l)$ are coordinates of 
the tangent vector to ${\cal M}_g$.

To fix a fundamental solution of (\ref{ls1}),(\ref{ls2}), impose the 
following normalization for some reference point $(z_0,\bz_0)\in\Si_g$ 
$$
\Psi(z_0,\bz_0)=I.
$$
Let $\ga$ be a homotopically nontrivial cycle in $\Si_g$ 
such that $(z_0,\bz_0)\in\ga$ and ${\cal Y}$ is the  corresponding 
monodromy transformation 
$$
{\cal Y}(\ga)=\Psi(z_0,\bz_0)|_\ga=P\exp\oint_\ga {\cal A} .
$$
The set of matrices $\{{\cal Y}(\ga)\}$ generates a representation of 
the fundamental group $\pi_1(\Si_g,z_0)$ in $\GLN$. 
Independence the monodromy ${\cal Y}$ on the deformations of the complex 
structure means that the linear equations
\beq{ls3}
\p_a{\cal Y}=0,~~(a=1,\ldots,l)~(\p_a=\p_{t_a}).
\eq
are consistent with (\ref{ls1}),(\ref{ls2}).
\begin{predl}
Equations (\ref{ls3}) are consistent with (\ref{ls1}),(\ref{ls2})
iff 
\beq{mot1}
\p_aA=0,~~(a=1,\ldots,l),
\eq
\beq{mot2}
\p_a\bA=\f1{\ka}A\mu_a^0,~~(a=1,\ldots,l).
\eq
\end{predl}
The proof is straightforward.
\begin{predl}
Equations of motion (\ref{mot1}),(\ref{mot2}) are Hamiltonian.
\end{predl}
Endow the space $FBun_{\Si,G}$ with the  
symplectic form
\beq{s0}
\om^{(0)}=\int_{\Si_g}<\de A,\de\bA>,~~(<,>=\tr),
\eq
and the set of Hamiltonians
\beq{ham}
H_a=\oh\int_{\Si_g}<A,A>\mu_a^{(0)},~~(a=1,\ldots,l).
\eq
Then (\ref{mot1}),(\ref{mot2}) are Hamiltonian equations with respect
to $\om^{(0)}$ and $H_s$.

Consider the bundle ${\cal P}$  
over the moduli space ${\cal M}_g$ with $FBun_{\Si,G}$ as the fibers.
The triple $(A,\bA,\bft)$ can be considered as the local coordinates of
the total space of the bundle. It is useful to consider ${\cal P}$
as the extended phase space \cite{Ar}.
There is a closed two-form on 
 ${\cal P}$
\beq{s1}
\om=\om^{(0)}-\f1{\ka}\sum_a\de H_a\de t_a.  
\eq
Though $\om$ is degenerated on ${\cal P}$ it produces 
the equations of motion (\ref{mot1}),(\ref{mot2}), since the form
$\om^{(0)}$ is non-degenerated along the fibers.

The gauge transformations in the deformed
complex structure take the form
\beq{gt}
A\rar f^{-1}\ka\p f+f^{-1}Af,~~
\bA\rar f^{-1}(\bp+\mu \p)f+ f^{-1}\bA f.
\eq
The form $\om$ is invariant under these transformations, though its 
constituents $\om^{(0)}$ and $H_s$ separately are not invariant.

Introduce a new couple of the connection components ${\cal A}=(A,\bA')$,
 where
$
\bA'=\bA-\f1{\ka}\mu A.
$
In terms of $(A,\bA')$ the form $\om$ (\ref{s1}) takes the canonical
form
\beq{s3}
\om=\int_{\Si_g}<\de A,\de\bA'>.
\eq
  
\bigskip
\noindent

{\bf 2.Symplectic reduction}\\
The form $\om$ is degenerated on  $FBun_{\Si,G}$, because 
it is invariant
under the action of the group ${\cal G}$ of gauge transformations 
(\ref{gt}), generating by the flatness condition (\ref{fl2}). 
The gauge fixing along with the flatness condition
(\ref{fl2}) is nothing else as the symplectic reduction from the space 
of smooth connections $Sm_{\Si,G}$ in the bundle $V_G$ 
to the reduced space
$$
\widetilde{FBun}_{\Si,G}=FBun_{\Si,G}/{\cal G}=Sm_{\Si,G}//{\cal G}
$$
The double slashes means that we impose the moment constraints
(\ref{fl2}) and fix the gauge.
$\widetilde{FBun}_{\Si,G}$ is {\sl the moduli space of flat connections}
 of the bundle $V_G$. In terms of 
the symplectic reduction procedure the flatness condition is called
{\sl the moment constraint equation}. 

Let us fix the gauge in a such way that the $\bA$ component
of ${\cal A}$ becomes anti-holomorphic 
\beq{gf}
\p\bar{L}=0,~~(\bar{L}= f^{-1}(\bp+\mu \p)f+ f^{-1}\bA f).
 \eq
We can do it because the antiholomorphity of 
$f^{-1}(\bp+\mu \p)f+ f^{-1}\bA f$
amounts to the classical equations of motion for the 
Wess-Zumino-Witten functional $S_{WZW}(f,\bA)$ for the gauge field $f$
in the external field $\bA$.  Denote the
gauge transformed field $A$ as $L$
$$
L=f^{-1}\ka\p f+f^{-1}Af.
$$
Then (\ref{fl2}) takes the form 
\beq{fl3}
(\bp +\p\mu)L+[\bar{L},L]=0.
\eq
Thus,  the moduli space of flat connections 
$\widetilde{FBun}_{\Si,G}$ are characterized 
by the set of solutions of the linear differential equation 
(\ref{fl3}) along with the condition (\ref{gf}).
The moduli space $\widetilde{FBun}_{\Si,G}$  
is finite-dimensional space
$$
\dim \widetilde{FBun}_{\Si,G}=2(N^2-1)(g-1),~~g>1.
$$
After the gauge fixing we come to the bundle $\ti{\cal P}$ over
${\cal M}_g$ with $\widetilde{FBun}_{\Si,G}$ as the fibers.
The system of linear differential equations (\ref{ls1}),(\ref{ls2})
and (\ref{ls3}) after the gauge fixing takes the form
\beq{ls4}
(\ka\p+L)\Psi=0,
\eq
\beq{ls5}
(\bp+\mu\p+\bar{L})\Psi=0,
\eq
\beq{ls6}
(\ka\p_s+M_s)\Psi=0,
\eq
where we replaced $\Psi$ on
$f^{-1}\Psi$ and $M_s=\ka\p_sff^{-1}$.

The gauge transformations do not spoil the consistency of the system. 
The consistency (\ref{ls4}) and
(\ref{ls5}) is provided by (\ref{fl3}) and (\ref{gf}). In fact,
the consistency (\ref{ls6}) with (\ref{ls4}) and (\ref{ls5})
leads to the Lax form of the equations of isomonodromic deformations
\beq{ide}
\p_sL-\ka\p M+[M,L]=0,
\eq
\beq{Me}
\ka\p_s\bar{L}-\mu^0_sL=(\bp+\mu\p)M_s-[M_s,\bar{L}].
\eq
They play the role (\ref{mot1}),(\ref{mot2}) correspondingly.
The last equation allows to find $M_s$ in terms of dynamical 
variables $L,\bar{L}$.

The symplectic 
form $\om$ on the reduced phase space   $\ti{\cal P}$ is
\beq{2.10}
\om=\int_{\Si_g}
<\de L,\de\bL>-\f1{\ka}\sum_s\de H_s\de t_s.
\eq
\beq{integr}
H_s=\oh \int_{\Si_g}<\de L,\de L>\mu_s^{(0)}
\eq

Introduce the local coordinates $(\bfv,\bfu)$ in 
$\widetilde{FBun}_{\Si,G}$:
$$
L=L(\bfv,\bfu,\bft),~\bar{L}=\bar{L}(\bfv,\bfu,\bft),
$$
$$
\bfv=(v_1,\ldots,v_{(N^2-1)(g-1)}),~~
\bfu=(u_1,\ldots,u_{(N^2-1)(g-1)}).
$$
Assume for simplicity that this parameterization leads to
the  canonical form on $\widetilde{FBun}_{\Si,G}$
\beq{ssf}
\om^{(0)}=\int_{\Si_g}<\de L(\bfv,\bfu,\bft),\de\bL(\bfv,\bfu,\bft)>=
(\de\bfv,\de\bfu),
\eq
where the pairing in the right hand side is induced by the trace.
The form on the extended phase space is
\beq{s4}
\om=(\de\bfv,\de\bfu)-\f1{\ka}\sum_s\de K_s(\bfv,\bfu,\bft)\de t_s,
\eq
and the variations of the Hamiltonians $K_s$ in the new variables 
take the form 
\beq{Ks}
\de K_s=
\int_{\Si_g}[<L,\de L>\mu_s^{(0)}+ \ka(<\de L,\p_s\bL>-<\p_s L,\de\bL>)]. 
\eq
Now, due to (\ref{fl3}), the hamiltonians depends explicitly 
on times.
Consider the one-form (the integral invariant
of Poincar\'{e}-Cartan)
$$
\te=\de^{-1}\om=
(\bfv,\de\bfu)-\f1{\ka}\sum_sK_s(\bfv,\bfu,\bft)\de t_s.
$$
There exist 
$3g-3=\dim{\cal M}_g$-dimensional space of vector fields 
${\cal V}_s$ that annihilates
$\te$ 
\beq{connec}
{\cal V}_s=\ka\p_s+\{H_s,\cdot\},~~(s=1,\ldots,l).
\eq
 It can be checked that ${\cal V}_s$ satisfy the following conditions
\beq{clflat}
 \ka\p_sH_r-\ka\p_rH_s+\{H_s,H_r\}_{\om^{(0)}}=0.
\eq
Thereby, they define  the flat connection in the bundle $\ti{\cal P}$.
These conditions are called {\sl the Whitham hierarchy} (WH).
The  equations for any function $f(\bfv,\bfu,\bft)$ on $\ti{\cal P}$ 
 take the form
\beq{eqm}
\frac{df(\bfv,\bfu,\bft)}{dt_s}=
\ka\frac{\p f(\bfv,\bfu,\bft)}{\p t_s}+\{H_s,f\}
\eq
They are called {\sl the hierarchy of isomonodromic deformations} (HID).

The both hierarchies can be derived from variations of the prepotential
${\cal F}$. It is defined 
 as the integral over the classical trajectories in
the extended phase space $\ti{\cal P}$ 
\beq{act}
{\cal F}(\bfu,\bft)={\cal F}(\bfu_0,\bft_0)+
\int_{\bfu_0,\bft_0}^{\bfu,\bft}
 {\cal L}_sdt_s,
\eq
where 
${\cal L}_s(\p_s\bfu,\bfu,\bft)=(\bfv,\p_s\bfu)-K_s(\bfv,\bfu,\bft),~
(\p_s\bfu=\frac{\de K_s}{\de\bfv})$ is the Lagrangian. ${\cal F}$
satisfies the set of the Hamilton-Jacobi equations
\beq{HJ}
\ka\p_s{\cal F}+H_s(\frac{\de {\cal F}}{\de \bfu},\bfu,\bft)=0.
\eq
The logarithm of ${\cal F}$ is called {\sl the tau-function}
of HID.
\bigskip
\noindent

{\bf 3. Singular curves} 

The singular curves are important for applications, since they
produce nontrivial systems for the low genus curves $(g=0,1)$.
In these cases the explicit calculations of hamiltonians
 are available. 

Consider a curve $\Si_{g,n}$ of genus $g$ with $n$
marked points $(x_1,\ldots,x_n)$. The number of times is equal to 
dimension of the moduli space 
${\cal M}_{g,n}$.  
We extend the space of connections $FBun_{\Si,G}=\{A,\bA\}$ 
by adding the coadoint orbits of $G=\GLN$ in the marked points
$$
({\cal O}_1,\ldots,{\cal O}_n),~~{\cal O}_b=\{p_b=gp_b^0g^{-1}\},
$$
where $p_b^0$ fixes the conjugacy class of ${\cal O}_b$.
We allow the $A$ component of connection to have simple 
 poles at the marked points, while the Beltrami differentials 
vanish there. Then instead of (\ref{fl2}) we obtain
\beq{s.1}
(\bp +\p\mu)A-\ka\p\bA+[\bA,A]=\sum_{b=1}^n\de^2(x_b)p_b
\eq
The Hamiltonian formalism is provided by the modified symplectic
form
\beq{s.2}
\om^{(0)}=\int_{\Si_{g,n}}<\de A,\de\bA>+
2\pi i\sum_{b=1}^n<\de (p_bg_b^{-1}),\de g_b>.
\eq
To derive HID one should start from the space of connections with
simple poles in the marked points.

Finally, we come to the same linear system (\ref{ls4}),
(\ref{ls5}),(\ref{ls6}), but due to (\ref{s.1}) the following 
relation between $L$ and $\bL$ holds
\beq{s.3}
(\bp +\p\mu)L+[\bar{L},L]=\sum_{b=1}^n\de^2(x_b)p_b.
\eq
As before, the linear equations are equivalent to the equations of
motion of HID coming from the symplectic form
\beq{s.4}
\om=\om^{(0)}({\bf v},{\bf u},{\bf p})-
\frac{1}{\ka}\sum_{s=1}^{l}
\de K_s({\bf v},{\bf u},{\bf p},{\bf t})\de t_s,~~(l=\dim({\cal M}_{g,n}),
\eq
where ${\bf p}=(p_1,\ldots,p_n)$,$\om_0$ is determined by the reduction from
(\ref{s.2})
\beq{s5}
\om^{(0)}=
\int_{\Si_{g,n}}
<\de L({\bf v},{\bf u},{\bf p}),\de\bL({\bf v},{\bf u},{\bf p})>
+2\pi i\sum_{b=1}^n<\de (p_bg_b^{-1}),\de g_b>,
\eq
 and $K_s$ (\ref{Ks}).

\section{Hitchin systems and their Whitham deformations}
\setcounter{equation}{0}
{\bf 1. Hitchin systems}\\
Consider the moduli space ${\cal R}_{g,N}$ of stable holomorphic 
$\GLN$ vector bundles $V$ over $\Si_g$. 
It is a smooth variety of dimension
\beq{4.1}
\dim {\cal R}_{g,N}=\ti{g}=N^2(g-1)+1.
\eq
Let $T^*{\cal R}_{g,N}$ be the cotangent bundle to  ${\cal R}_{g,N}$
with the standard symplectic form on it. 
Hitchin \cite{H} defined a completely integrable system on 
$T^*{\cal R}_{g,N}$. 

The space $T^*{\cal R}_{g,N}$ can be obtained by the symplectic
reduction from the space 
$T^*{\cal R}_{g,N}^s=(\Phi,\bA)$, where 
$\bA$ is a smooth connection of the stable bundle corresponding to 
$\bp+\bA$ and $\Phi$ is the {\sl Higgs field}
$\Phi\in\Om^0(\Si_g,End V\otimes K)$ 
($K$ is the canonical bundle of $\Si_g$). There is the well defined
symplectic form on this space
\beq{4.2}
\om^{(0)}=\int_{\Si_g}<\de\Phi,\de\bA>.
\eq
This form is invariant with respect to the gauge group
 ${\cal G}=C^\infty Map(\Si_g,\GLN)$ action
\beq{4.3}
\Phi\rar f^{-1}\Phi f,~~\bA\rar f^{-1}\bp f+f^{-1}\bA f.
\eq
In particular,
$
{\cal R}_{g,N}={\cal R}_{g,N}^s/{\cal G}.
$
Let $\rho_{s,k}=\rho_{s,k}\p_z^{k-1}\otimes d\bz$ be the 
$(-k+1,1)$-differentials 
($\rho_{s,k}\in H^1(\Si_g, \G^{k-1}\otimes K)$), and $s$ 
enumerates the basis in $ H^1(\Si_g, \G^{k-1}\otimes K)$.  
 ($\rho_{s,2}=\mu_s$).
Due to the Riemann-Roch theorem 
$$
\dim H^1(\Si_g, \G^{k-1}\otimes K)=(2k-1)(g-1). 
$$
These differentials allows to define the gauge invariant Hamiltonians
\beq{4.4}
H_{s,k}=\f1{k}\int_{\Si_g}<\Phi^k>\rho_{s,k},~~
(k=1,\ldots,N,~s=1,\ldots,(2k-1)(g-1)).
\eq
The Hamiltonian equations take the form
\beq{4.5}
\p_a\Phi=0,~~(\p_a=\frac{\p}{\p t_a}, a=(s,k)),
\eq
\beq{4.6}
\p_a\bA=\Phi^{k-1}\rho_{s,k}
\eq

The gauge action produces the moment map 
$
\mu: ~T^*{\cal R}^s\rar Lie^*(\GLN).
$
It follows from (\ref{4.2}),(\ref{4.3})  that
$
\mu=\bp\Phi+[\bA,\Phi].
$
The reduced phase space is the cotangent bundle we started with
$$
T^*{\cal R}_{g,N}\sim T^*{\cal R}^s//{\cal G}:=\mu^{-1}(0)/{\cal G}.
$$
The {\sl Hitchin hierarchy} (HH)  is the set of Hamiltonian equations with
$H_{s,k}$ (\ref{4.4}) on the reduced phase space $T^*{\cal R}_{g,N}$.
Hitchin observed that the number of  integrals $H_{s,k}$
$$
\sum_{k=1}^N(2k-1)(g-1)=N^2(g-1)+1
$$
coincides with dimension $\ti{g}$ of the coordinate space ${\cal R}_{g,N}$
(\ref{4.1}). Since they are independent and Poisson-commute, 
HH is the set of completely integrable Hamiltonian systems.

Let fix the gauge of the field $\bA$
$$
\bA=f\bp f^{-1}+f\bL f^{-1}.
$$
Then 
$$
L=f^{-1}\Phi f.
$$
is a solution of the moment constraint equation
\beq{4.7}
\bp L+[\bL,L]=0.
\eq
The space of solutions of this equation is isomorphic to 
$H^0(\Si_g,End ~V\otimes K)$ - the cotangent space to the moduli
space ${\cal R}_{g,N}$.

The gauge transformation $f$ defines the element $M_a$ in Lie$(\GLN)$ 
$M_a=\p_aff^{-1}$.
\begin{predl}
The system of linear equations
\beq{4.9}
(\la+L)Y=0,
\eq
\beq{4.10}
(\bp+\sum_{s,k}\la^{k-1}t_{s,k}\rho_{s,k}+\bL)Y=0,
\eq
\beq{4.11}
(\p_a+M_a)Y=0
\eq
is consistent and defines the equations of motion for HH.
\end{predl}
{\sl Proof}. The consistency of (\ref{4.9}) and (\ref{4.10}) follows 
from (\ref{4.7}).
In terms of $L$ the equations of motion (\ref{4.5}),(\ref{4.6}) 
take the form
\beq{4.12}
\p_aL+[M_a,L]=0 ~({\rm the~Lax~equation}),
\eq
\beq{4.13}
\p_a\bL-\bp M_a+[M_a,\bL]=L^{k-1}\rho_{s,k},~~(a=(s,k)).
\eq
The Lax equation provides the consistency of (\ref{4.9}) and (\ref{4.11}),
while the second equation of motion (\ref{4.13}) plays the same role for 
the couple (\ref{4.10}) and (\ref{4.11}). 
This equation allows to determine $M_a$ from $L$ and $\bL$.

The bundles over singular curves can be incorporated in this approach
as well \cite{Ne}. The Higgs field has simple poles at the marked points
and (\ref{4.7}) is replaced by 
\beq{4.14}
\bp L+[\bL,L]=2\pi i\sum_{b=1}^n\de^2(x_b)p_b.
\eq
The form $\om^{(0)}$ on $T^*{\cal R}_{g,N}$
\beq{4.30} 
\om^{(0)}=\int_{\Si_g}<\de L,\de\bL>
+2\pi i\sum_{b=1}^n<\de (p_bg_b^{-1}),\de g_b>
\eq
Note that the space $T^*{\cal R}_{g,N}$ and
the space of flat bundles $\widetilde{FBun}_{\Si,G}$ have the same 
dimensions. Moreover, it follows from the comparison 
the moment constraints (\ref{s.3}),(\ref{4.14}) and the symplectic 
forms (\ref{s5}),(\ref{4.30}) that
 they are isomorphic as symplectic manifolds.

\bigskip
\noindent

{\bf 2.Spectral description}\\

Due to the Liouville theorem  the phase flows of HH are restricted to
the Abelian varieties, corresponding to a level set of 
the Hamiltonians $H_{s,k}=c_{s,k}$. 
The phase flow takes a simple form in terms of 
action-angle coordinates. They are defined in a such way that
the angle type coordinates are angular coordinates on the
 Abelian variety,  and the hamiltonians depend on the action
coordinates only.
 To describe them
consider the characteristic polynomial of the matrix $L$ 
\beq{4.15}
P(\la,z)=\det(\la+L)=\la^N+b_1\la^{N-1}+\ldots+b_j\la^{N-j}+\ldots+b_N,
\eq
$$
b_j=\sum Min_j, ~~( Min_j- {\rm principle ~minors~of~order~}j,~ 
b_1=\tr L,~b_N=\det L).
$$

The spectral curve ${\cal C}\subset T^*\Si_g$ is defined as the 
zero set of $P$
$$
C=\{P(\la,z)=0\}.
$$
It is a well defined object, because the coefficients $b_j$ are
gauge invariant.

Since $L\in H^0(\Si_g,End ~V\otimes K)$, the coefficients
$b_j\in H^0(\Si_g, K^j)$ and we obtain the map
\beq{4.17}
p:T^*{\cal R}_{g,N}\rar B=\oplus_{j=1}^NH^0(\Si_g, K^j).
\eq
The space $B$ can be considered as the moduli space of the family
of spectral curves parameterized by the Hamiltonians $H_{s,k}$. 
The fibers of $p$ are Lagrangian subvarieties of $T^*{\cal R}_{g,N}$.
The spectral curve ${\cal C}$ is the $N$-fold covering of the basis curve 
$\Si_{g,N}$
$$
\pi:C\rar\Si_{g,N}.
$$
 Its genus $g({\cal C})$ is equal to dimension $\ti{g}$ of 
${\cal R}_{g,N}$. There is a line bundle ${\cal L}$ with
an eigenspace of $L(z)$ 
corresponding to the eigenvalue $\la$ as a fiber
over a generic point $(\la,z)$ 
$$
{\cal L}\subset\ker(\la+L)\subset\pi^*(V).
$$
It defines a point of the Jacobian $Jac({\cal C})$, the Liouvillean
variety of dimension $\ti{g}=g({\cal C})$.

 Conversely, if $z\in\Si_{g}$ is not a branch point one can 
reconstruct $V$ for a given line bundle on ${\cal C}$ as
$$
V_z=\oplus_{v\in\pi^{-1}(z)}{\cal L}_v.
$$

Let $\om_j,~j=1\ldots,\ti{g}$ be the canonical holomorphic 
one-differentials on ${\cal C}$ such that for the cycles
$\al_1,\ldots,\al_{\ti{g}};\be_1,\ldots,\be_{\ti{g}},~~
\al_i\cdot\al_j=\be_i\cdot\be_j=0,~~
\al_i\cdot\be_j=\de_{ij},~~
\oint_{\al_i}\om_j=\de_{ij}$.
Then the symplectic form $\om^{(0)}$ (\ref{4.2})
can be written in the form
$$
\om^{(0)}=\int_{\Si_g}<\de L,\de\bL>=
\sum_{j=1}^N\int_{\Si_g}\de\la_j\de\xi_j.
$$
Here $\xi_j$ are diagonal elements of $s\bL s^{-1}$,
where $s$ diagonalizes 
$L,~sLs^{-1}=\di(\la_1,\ldots,\la_N)$.
Then we obtain
$$
\om^{(0)}=\int_{{\cal C}}\de\la\de\xi.
$$ 
Because $\la$ is a holomorphic one-form on ${\cal C}$, it can be
decomposed as 
$
\la=\sum_{j=1}^{\ti{g}}a_j\om_j.
$
Thereby
$$
\om^{(0)}=\sum_{j=1}^{\ti{g}}\de a_j\int_{{\cal C}}\om_j\de\xi.
$$
The action variables can be identify with
\beq{4.19a} 
 a_j=\oint_{\al_i}\la\om_j,~~(j=1\ldots,\ti{g}).
\eq
To define the angle variables, put locally $\xi=\bp\log\psi$. 
If $(p_m)$ is a divisor of $\psi$ then 
$$
\int_{{\cal C}}\om_j\de\xi=\sum_m\int_{p_0}^{p_m}\om_j\log\psi
=\de\vf_j.
$$
Thus $\vf_j$ are linear coordinates on $Jac({\cal C})$ and
$$
\om^{(0)}=\sum_{j=1}^{\ti{g}}\de a_j\de\vf_j.
$$

\bigskip
\noindent

{\bf 3. Scaling limit.}

Consider HID in the limit $\ka\rar 0$. 
The value $\ka=0$ is called critical.
We prove that on the critical level HID coincide with part of
HH relating to the quadratic Hamiltonians (\ref{4.4}).

Note first, that  in this limit the $A$-connection is
transformed in the Higgs field $\Phi$: 
$
A\stackrel{\ka\rar 0}\longrightarrow\Phi,
$
and therefore
$$
\widetilde{FBun}_{\Si,G}\stackrel{\ka\rar 0}\longrightarrow
 T^*{\cal R}_{g,N}.
$$
But the form $\om$ on the extended phase ${\cal P}$ appears to be 
singular (see (\ref{s1}),(\ref{2.10})).
 To get around we rescale the times
\beq{4.20}
{\bf t}=\bf T +\ka{\bf t}^H,
\eq
where  ${\bf t}^H$ are the fast (Hitchin) times and $\bf T$ are the
slow times. Assume that only fast times are dynamical. It means that
$$
\de\mu({\bf t})=\ka\sum_s\mu_s^{(0)}\de t_s^H,~~
(\mu_s^{(0)}=\bp n_s).
$$
After this rescaling the forms  (\ref{s1}),(\ref{2.10})
 become regular.
The rescaling procedure means that we blow up a vicinity
 of the fixed point $\mu_s^{(0)}$
 in ${\cal M}_{g,n}$ and the whole dynamic 
 is developed in this vicinity.
This fixed point is defined by the complex coordinates
\beq{fp}
w_0=z-\sum_sT_s\ep_s(z,\bz),~~\bar{w}_0=\bz.
\eq

Now compare the Baker-Akhiezer function of HID $\Psi$ (\ref{ls4}),
(\ref{ls5}),(\ref{ls6}) with the Baker-Akhiezer function of HH $Y$ 
(\ref{4.9}),(\ref{4.10}),(\ref{4.11}).
Using the WKB approximation, assume that 
\beq{WKB}
\Psi=\Phi\exp(\frac{{\cal S}^{(0)}}{\ka}+{\cal S}^{(1)}),
\eq
where $\Phi$ is a  group valued function and ${\cal S}^{(0)}$,
${\cal S}^{(1)}$ are diagonal matrices. Let substitute (\ref{WKB})
 in the linear system (\ref{ls4}),(\ref{ls5}),(\ref{ls6}). If
$$
\frac{\p}{\p\bar w_0}{\cal S}^{(0)}=0,~~
\frac{\p}{\p t^H_s}{\cal S}^{(0)}=0.
$$
there are no terms of order $\ka^{-1}$.
It follows from the definition of the fixed point in the
moduli of complex structures (\ref{fp}) that
\beq{S0}
{\cal S}^{(0)}={\cal S}^{(0)}(T_1,\ldots,T_l|z-\sum_sT_s\ep_s(z,\bz)).
\eq
We take also
$
{\cal S}^{(1)}=\p{\cal S}^{(0)}\sum_s t^H_s\ep_s(z,\bz).
$
In the quasi-classical limit we put
 \beq{SW}
 \p{\cal S}^{(0)}=\la.
\eq

In the zero order approximation we come to the linear system 
of HH (\ref{ls4}),(\ref{ls5}),(\ref{ls6}), defining by the 
Hamiltonians $H_{k,s},~k=1,2$.
The Baker-Akhiezer function $Y$ takes the form
\beq{Y}
Y=\Phi e^
{
\sum_s t_s^H
\frac{\p}{\p T_s}{\cal S}^{(0)}
}.
\eq

Our goal is the inverse problem. 
We need to reconstruct the dependence on the slow times ${\bf T}$
starting from solutions of HH. Since ${\bf T}$ is a vector
in the tangent space to the moduli of curves ${\cal M}_{g,n}$,
it defines a deformation of the spectral curve in the space
$B$ (see (\ref{4.4}),(\ref{4.17})).
Solutions $Y$ of the linear systems (\ref{4.9})(\ref{4.10}),(\ref{4.11})
take the form $Y=\Phi e^{\sum_s t_s^H\Om_s}$, where $\Om_s$ are
diagonal matrices. Their entries are primitive functions of meromorphic
differentials with singularities matching the corresponding poles of $L$.
Then according with (\ref{Y}) we can assume that
$$
\frac{\p}{\p T_s}d{\cal S}=d\Om_s.
$$
 These equation define the approximation to the
phase of $\Psi$ in the linear problems (\ref{ls4}),(\ref{ls5}),\\
(\ref{ls6}) 
of HID along with
 $$
\frac{\p}{\p a_j}d{\cal S}=\om_j.
$$
The differential $d{\cal S}$ plays role of the Seiberg-Witten differential.
Important point is that only part of the spectral moduli, connected with
$H_{k,s},~k=1,2$, is deformed. As a result there is no matching between
the action parameters of the spectral curve $a_j,j=1,\ldots,\ti{g}$
(\ref{4.19a}) and deformed hamiltonians.
The detailed analyses of this situation in the 
rational case is undertaken in \cite{Ta}.

Another object of the Whitham quantization is the prepotential
${\cal F}$ (\ref{act}). It depends on the action variables 
$a_j$ . This dependence is compatible with
 the Hamilton-Jacobi equation (\ref{HJ}) with slow times $T_s$ as 
the independent variables. These equations are discussed  in
\cite{Ta,IM}.

\section{Classical limit of
the Knizhnik-Zamolodchikov-Bernard equations}
\setcounter{equation}{0}

The Knizhnik-Zamolodchikov-Bernard equations (KZB) are the system
of differential equations  having the form  
of the non-stationer Schr\"{o}dinger equations with the times 
coming from ${\cal M}_{g,n}$ (see, for example, \cite{I2}).

 They arise in the geometric quantization of the
moduli of flat bundles $\widetilde{FBun}_{\Si,G}$ \cite{ADPW,H2}.
Let $V=V_1\times\cdots\otimes V_n$ be the 
tensor product of finite-dimensional
irreducible representations associated with the marked points.
The Hilbert space  of the quantum system is a space of sections 
of the bundle ${\cal E}_{V,\ka^{quant}}(\Si_{g,n})$   over 
$\widetilde{FBun}_{\Si,G}$ depending on
an negative number $\ka^{quant}$ with the $V$-fibers. 
It is the space of conformal blocks of the
WZW theory on $\Si_{g,n}$.

The Hitchin systems are the classical limit of the KZB equations
on the critical level \cite{Ne,I2}.
  The classical limit means that one replaces operators
by their symbols
and generators of finite-dimensional representations in the vertex
 operator acting in the spaces $V_j$ by the corresponding elements 
of coadjoint orbits.
To pass to the classical limit in the KZB equations 
\beq{NS}
(\ka^{quant}\p_s+\hat {H}_s)F=0.
\eq
we replace the conformal block by its quasi-classical expression
\beq{cb}
F=\exp \frac{\cal F}{\hbar},
\eq
where $\hbar =(\ka^{quant})^{-1}$. Consider the classical limit
$  \ka^{quant} \rar\infty$ and assume that values of the Casimirs 
$C^i_a,~(i=1,\ldots,{\rm rank}G,~a=1,\ldots,n)$ 
corresponding to the
irreducible representations defining the vertex operators 
 also go to infinity. Let all values 
$\lim\frac{C_a^i}{\ka^{quant}}$
are finite. It allows to fix the coadjoint orbits in the marked points.
In the classical limit (\ref{NS}) is transformed to the Hamilton-Jacobi
equation for the action ${\cal F}=\log\tau$ (\ref{HJ}) of HID. 

The integral representations of conformal blocks
are known for WZW theories over rational and elliptic curves 
\cite{SV,FG,EK,FW}. 
Then (\ref{cb}) allows to extract the prepotential ${\cal F}$ of HID.  

The KZB operators (\ref{NS}) play role of flat connections in the bundle
${\cal P}^{quant}$ over the moduli of curves ${\cal M}_{g,n}$ with
the fibers ${\cal E}_{V,\ka^{quant}}(\Si_{g,n})$ \cite{Hi2,Fe}
$$ 
[\ka^{quant}\p_s+\hat {H}_s,\ka^{quant}\p_r+\hat {H}_r]=0.
$$
These equations is the quantum counterpart of the Whitham hierarchy
(\ref{clflat}).

\section{Multicomponent generalization of PVI$_\nu$}
\setcounter{equation}{0}

Consider $\widetilde{FBun}_{\Si,G}$ over the family of elliptic
curves 
 with a one marked point ${\cal M}_{1,1}$. The space ${\cal M}_{1,1}$
is one-dimensional, because the position
of one point on a torus is irrelevant. Thus, 
we have only one time $\tau$ and ${\cal M}_{1,1}\sim E_\tau$
(\ref{ef}).
In this case the Beltrami differential takes the form
$$
\mu=\frac{\tau-\tau_0}{\tau-\bar{\tau}_0}.
$$

Consider the most degenerated orbit ${\cal O}=(gp^0g^{-1})$ 
of $\GLN$ sitting in the marked point $z=0$ with
\beq{6.1}
p^0=\nu[(\underbrace{1,\ldots,1}_N)^T\otimes
(\underbrace{1,\dots,1}_N)-Id].
\eq
For stable bundles the gauge transforms 
allow to put $\bA$ component in the diagonal form
\beq{6.2}
\bar{L}= \frac{2\pi i}{\tau-\bar{\tau}_0}{\bf u},
~~{\bf u}=\di(u_1,\ldots,u_N)\in{\cal H}-{\rm Cartan~ algebra}.
\eq
It means that
\beq{6.3}
\int_{E_\tau}\bar{L}dwd\bar{w}={\bf u}.
\eq
Let $\bar{L}=\bp\log {\bf \phi}$. Then the integral
$$
\int_{E_\tau}\bar{L}dwd\bar{w}=\int_{P_0}^P\log {\bf \phi}dw.
$$
defines the Abel map $E_\tau$ in the  product of $N$
Jacobians.

The remaining gauge transforms do not change the gauge fixing.
These transformations are generated by the Weyl
 subgroup $W$ of $G$ and  elements
\mbox{$f(w,\bar{w})\in {\rm Map}(T^2_{\tau},{\rm Cartan}(G))$}.
The orbit variables can be gauged away by these transforms 
and we are left with $p^{(0)}$ (\ref{6.1}).
The solution $L$ of the moment constraint
$$
\p_{\bar{w}}L+[\bar{L},L]=2\pi i\de^2(0)p^{(0)}
$$
takes the form
\beq{Lop}
L=P+X,~
P=2\pi i(\frac{{\bf v}}{1-\mu}-\ka\frac{{\bf u}}{\rho}),
\eq
$$
{\bf u}=\di(u_1,\ldots,u_N),~~{\bf v}=\di(v_1,\ldots,v_N)
$$
$$
X_{jk}=x(u_j-u_k)=(\tau-\bar{\tau}_0)\nu\exp 2\pi i\{
\frac{w-\bar{w}}{\tau-\bar{\tau}_0}
(u_j-u_k)\}
\phi(\al(u_j-u_k),w),
$$
$$
\phi(u,z)=
\frac
{\te(u+z)\te'(0)}
{\te(u)\te(z)},~~
\te(z|\tau)=q^{\frac
{1}{8}}\sum_{n\in {\bf Z}}(-1)^ne^{\pi i(n(n+1)\tau+2nz)}.
$$
The operator $M$ defining the phase flow according with the Lax
equation (\ref{ide}) can be extracted from (\ref{Me})
\beq{Mop}
M=-D+Y,~D=\di(d_1,\ldots,d_N),~~d_j=\sum_{i\neq j}^Ns(u_j-u_i),~
s(u)=\f1{\ka}\wp(u)+const.
\eq
$$
Y_{jk}= y(u_j-u_k),~~y(u,w,\bar{w})=
\frac{\rho}{2\pi i\ka(\tau-\bar{\tau}_0)}\p_u x(u,w,\bar{w}).
$$
The functions $x,y,z$ satisfy the functional equations
\beq{fe}
x(u,z,\bz)y(v,z,\bz)-x(v,z,\bz)y(u,z,\bz)=(s(v)-s(u))x(u+v,z,\bz).
\eq
This equation is derived from the Lax equation (\ref{ide}).

The symplectic form $\om$ is boiled down to
$$
\om=(\de \bfv,\de\bfu)-\f1{\kappa}\de H\de\tau,
$$
where 
$$
H=\frac{(\de \bfv,\de \bfv)}{2}-\frac{\nu^2}{(2\pi i)^2}
\sum_{j<k}^N\wp(u_j-u_k|\tau).
$$
They define the hamiltonian flow
\beq{6.4}
\frac{d^2u_j}{d\tau^2}=\frac{\nu^2}{(2\pi i)^2}
\sum_{k< j}^N\wp_u(u_j-u_k|\tau).
\eq
For$N=2$ one can put $u_1=-u_2=u$. Then the potential
$$
\frac{\nu^2}{(2\pi i)^2}\wp(2u|\tau)=\frac{\nu^2}{(2\pi i)^2}
\sum_{j=0}^3\wp(u+\frac{T_j}{2}|\tau)
$$
produces PVI$_\nu$ (see (\ref{I.3}).
 
The remaining gauge symmetries implies that $\om$ is invariant
under the Weyl transformations $W$ of $(\bfv,\bfu)$ and the lattice
actions (compare with (\ref{1.3})
$$
\bfv\rar s\bfv,~\bfv+\ka{\bf n},~
\bfu\rar s\bfu,~\bfu-{\bf m}+\tau{\bf n},
 ~(s\in W,~{\bf n}\in{\bf Z}^N).
$$ 
It is also invariant under the $\PSL$ action on $\tau$
$$
\tau\rar\frac{a\tau+b}{c\tau+d},
~
\bfv\rar \bfv(c\tau+d)-\ka c\bfu,
~
\bfu\rar\bfu (c\tau+d)^{-1}.
$$
This invariance is follows from the invariance of the upstairs
system under the diffeomorphisms of $\Si_{g,n}$ (see \cite{LO} 
for detailes). 

On the critical level $\ka\rar 0,~\tau-\tau_0=\ka t$ we 
obtain the elliptic Calogero N-body system. This system is a
particular example of  the Hitchin systems \cite{Ne}.
Note, that the functions $x,y$ and $s$ defining the Lax matricies
 satisfy the same functional
equation (\ref{fe}) as in the Calogero-Hitchin limit $\ka=0$
\cite{Ca}.

\section{Conclusion}
Here we propose a few open problems in the context of
 topics discussed above.

\bigskip
\noindent

$\bullet $
The evident problem is a description of PVI with four arbitrary constants 
as a reduced  Hamiltonian system. The first step in this direction 
is the Lax form of
$PVI_{\al,\be,\ga,\de}$. The Lax form is even unknown on the critical level,
i.e. for the Calogero-Inozemtsev system. It will be interesting
to generalize this approach to the $N$-body Calogero-Inozemtsev system and
the $N$-component PVI with four coupling constants.

\bigskip
\noindent

$\bullet$ 
The degenerations of PVI to PV-PI in terms of elliptic functions.

\bigskip
\noindent

$\bullet$ 
There exists a generalization of the Calogero systems related to any simple
group. In addition to degrees of freedom coming from the moduli of bundles
(the coordinates of particles),
these systems certainly contain degrees of freedom related to the coadjoint 
orbits. Recently a new Lax equations based on
arbitrary root systems without the orbit coordinates were proposed
\cite{Ph,Sa}. This construction is purely algebraic and does not use
 the symplectic reduction. How these systems can be incorporated
 in the Hitchin approach, 
or, more generally, in the isomonodromic deformation construction?

\bigskip
\noindent

$\bullet$ 
Consider the $N=2$ elliptic Calogero system. The solution $u(t)$, 
corresponding to the fixed value $h_2$ of the Hamiltonian
$$
H=\frac{v^2}{2}+\frac{\nu^2}{4\pi^2}\wp(2u|\tau_0)=h_2,
$$
is implicitly described by the elliptic integral of the first kind
$$
t-t_0=\oh\int_{2u_0}^{2u}\frac{dx}{y'},~~y'=y\sqrt{2h_2-\frac{\nu^2}{2\pi^2}},
$$
where $y=4(x-e_1(\tau_0))(x-e_2(\tau_0))(x-e_3(\tau_0))$. As it was mentioned
 at the end of Sect.6  it can serve for the  calculations of solutions
to PVI$_\nu$. This procedure can be accomplished by the Krichever averaging
method \cite{Kr3}. It will be interesting to compare this approximation
with explicit solutions presented recently in \cite{It,Ko} for some particular 
value of the coupling constant $\nu$.

As suggested in Sect.5 another way of approximation  
 comes from the classical limit 
of conformal blocks for $SL_2({\bf C})$ theory on elliptic curves
with one marked point \cite{EK,FV}. Which method gives the better
approximation?

\bigskip
\noindent

$\bullet$
We considered deformations with respect to the
moduli of complex structures of curves. They describe only part of the moduli
of the spectral curves ${\cal C}$. The remaining moduli of ${\cal C}$ come
from $\rho_{s,k},~k>2$. They correspond to the so-called W-geometry of the
 basic curve $\Si_{g,n}$. This geometry
is poorly understood. On the other side, there are no examples of 
isomonodromic deformation equations with respect to these moduli spaces, 
as well as the corresponding higher order KZB equations. 
Any progress in understanding of one of these subjects will shed light on 
another.

{\bf Acknowledgments}\\
{\sl The work is
supported in part by grants RFFI-96-02-18046 INTAS 96-518 and
 96-15-96455 for support of scientific schools. 
I am grateful to the Max-Planck-Institut f\"{u}r Mathemamatik in Bonn
for the hospitality, where this paper was prepared.
}

\small{

}
\end{document}